\documentclass[aps,prd,
floatfix,preprintnumbers,superscriptaddress,nofootinbib,nameinlink,capitalise]{revtex4-2}

\usepackage[left=2.54cm, right=2.54cm, top=2.54cm, bottom=2.54cm]{geometry}
\usepackage{palatino}
\usepackage{amsmath, amssymb}
\usepackage{graphicx}
\usepackage{array, booktabs, multirow}
\usepackage{hyperref}
\usepackage{subfig}
\usepackage{nameref}
\usepackage{cleveref}
\usepackage{placeins}
\usepackage{verbatim}
\usepackage{tikz}
\usepackage{array}
\usepackage{rotating}

\usepackage{tabularx}
\usepackage[ruled,vlined]{algorithm2e}
\DontPrintSemicolon
\SetKwComment{Comment}{$\triangleright$\ }{}
\SetKwInput{KwIn}{Input}
\SetKwInput{KwOut}{Output}

\usepackage{cleveref}  
\usepackage{float}

\usetikzlibrary{shapes.geometric, arrows}

\tikzstyle{startstop} = [rectangle, rounded corners, minimum width=3cm, minimum height=1cm,text centered, draw=black, fill=red!30]
\tikzstyle{process} = [rectangle, minimum width=3cm, minimum height=1cm, text centered, draw=black, fill=blue!20]
\tikzstyle{decision} = [diamond, minimum width=3cm, minimum height=1cm, text centered, draw=black, fill=green!30]
\tikzstyle{arrow} = [thick,->,>=stealth]

\usepackage{textgreek}
\usepackage[justification=raggedright,singlelinecheck=false]{caption}
\usepackage{color}
\usepackage{xcolor}

\newcommand{\bmat}{\left(\begin{array}}
\newcommand{\emat}{\end{array}\right)}
\newcommand{\be}{\begin{equation}}
\newcommand{\ee}{\end{equation}}
\newcommand{\bea}{\begin{eqnarray}}
\newcommand{\eea}{\end{eqnarray}}

\usepackage{graphicx}
\usepackage{subcaption}

\begin{document}

\title{Quantum AI for Cybersecurity: A hybrid Quantum-Classical models for attack path analysis }

\author{Jessica A. Sciammarelli}

\author{Waqas Ahmed}
\email{waqasmit@hbpu.edu.cn}
\affiliation{Center for Fundamental Physics, School of Artificial Intelligence, Hubei Polytechnic University, Huangshi 435003, China}
\affiliation{The European Higher Education Institute,
St. Julian's STJ 3141, Malta}
\email{Jessicaengenhariabr@hotmail.com}


\noaffiliation


\vspace{1em}
\begin{abstract}
Modern cyberattacks are increasingly complex, posing significant challenges to classical machine learning methods, particularly when labeled data is limited and feature interactions are highly non-linear. This study investigates the potential of hybrid quantum-classical learning to enhance feature representations for intrusion detection and explore possible quantum advantages in cybersecurity analytics. Using the UNSW-NB15 dataset, network traffic is transformed into structured feature vectors through classical preprocessing and normalization. Classical models, including Logistic Regression and Support Vector Machines with linear and RBF kernels, are evaluated on the full dataset to establish baseline performance under large-sample conditions. Simultaneously, a quantum-enhanced pipeline maps classical features into variational quantum circuits via angle encoding and entangling layers, executed on a CPU-based quantum simulator, with resulting quantum embeddings classified using a classical SVM. Experiments show that while classical models achieve higher overall accuracy with large datasets, quantum-enhanced representations demonstrate superior attack recall and improved class separability when data is scarce, suggesting that quantum feature spaces capture complex correlations inaccessible to shallow classical models. These results highlight the potential of quantum embeddings to improve generalization and representation quality in cybersecurity tasks and provide a reproducible framework for evaluating quantum advantages as quantum hardware and simulators continue to advance.

\end{abstract}

\maketitle

\vspace{5.9cm}
\textbf{Keywords:} Cybersecurity, Quantum Artificial Intelligence, Hybrid Models, Attack Path Analysis, Quantum Machine Learning, Graph Learning, Quantum Feature Spaces, Variational Quantum Circuits

\newpage
\section{Introduction}\label{sec:introduction}

The modern-day cyber threat environment is marked by the rapid development and growing complexity, posing an uphill challenge to defensive security paradigms. Contemporary organizational networks are extremely interdependent systems, with one vulnerability spreading to various channels, forming complicated attack chains that are hard to predict and eliminate with the help of traditional analytical software \cite{CHAWLA20232191}. Attack graph analysis has emerged as a fundamental method of modeling such possible exploit sequences, and networks states and transitions are modeled as graphical structures where nodes represent system components and edges potential adversarial actions\cite{APOLLONI1989233}. Classical computational methods of graph security analysis impose an inherent limitation on the complexity of modern attack surfaces, and insidious and dynamic trends of advanced persistent threats, despite their remarkable progress.

Quantum computing brings a paradigm change in the theory of computation, promising to provide a yield of the speed up, of specific types of problems, which are considered to be intractable to classical computing \cite{Pre}. In cybersecurity, quantum tools are being pursued to find a variety of applications, such as cryptographic analysis, encrypted traffic inspection, and improved intrusion detection systems \cite{Movassagh}. Quantum machine learning (QML), which takes advantage of special quantum-mechanical effects such as superposition and entanglement, can possibly support the processing of information in radically new ways \cite{Biamonte2017QML}. Early studies indicate promising results, such as quantum support vector machines achieving high accuracy in malware classification \cite{Xiong:2024gld} and quantum ensemble models effectively identifying malicious code \cite{Wu_2021}. However, the   research on quantum-enhanced learning applied to attack path analysis remains
largely unexplored, so a significant research gap exists in the area intersection of quantum computation and cybersecurity analytics.

The paper fills a major research gap at the interface of quantum computation and cybersecurity analytics the formal use of quantum-enhanced learning in the context of attack path analysis.. It is observed that classical graph learning algorithms are practical but fail to handle the complex nonlinear pattern of relationships, and exponentiation of the size of the attack graph states of a real attack attacker defender image \cite{Fahim}. The concept of quantum feature spaces capable of modeling information on exponentially dimensional Hilbert spaces is a theoretical concept that makes it possible to describe such complex patterns in a more efficient way \cite{Schuld:2018uel}. However, they have not been empirically validated to allow their practical use in the resource-constrained environment, and particularly, in the settings that can be reached by the broader research community.

The primary research question that will be asked in this study is:

\textbf{Can hybrid quantum-classical models demonstrably outperform purely classical approaches for attack path analysis, especially when training data is limited?}
Our hypothesis is that quantum feature embeddings have the capability of high inter-class separabilities and better generalization when applied to attack graph classification tasks and under data-poor environments like those observable in cybersecurity processes in reality.

To investigate this hypothesis, we create and execute a five-step hybrid architecture which is evaluated on the \textbf{UNSW-NB15 dataset} \cite{moustafa2015unsw}  a widely adopted benchmark for network intrusion detection containing approximately 2.5 million network flows with both normal and attack traffic. Our methodology integrates classical graph preprocessing with quantum feature mapping, employing angle encoding and variational quantum circuits (VQCs) to transform classical features into quantum representations. These quantum embeddings are then combined with classical machine learning classifiers (Support Vector Machines) for attack classification, with comparative evaluation against classical baselines (Logistic Regression, Linear SVM, RBF SVM) under both data-rich and data scarce conditions.

We find that, as long as classical models would be relatively robust to large amounts of training data (with the accuracy of those models being around 69\% on the full UNSW-NB15 dataset), quantum embeddings demonstrate promising sensitivity to attack patterns even with severely limited samples (200 instances). Specifically, quantum-enhanced models had perfect recall (100\%) with attack classes in low-data conditions, though with trade-offs in benign class detection. These results indicate that quantum feature spaces have representational benefits on attack-sensitive tasks, as well as indicate the existing limitations in circuit expressivity and class balance that need additional optimisation.

The key contributions of this work are summarized as follows:

\begin{itemize}
    \item An innovative and scalable scheme of encoding graph-based cybersecurity data into quantum representations to analyze a path of attack..
    \item Empirical comparison between quantum graph kernels and variational quantum circuits on attack graph classification..
    \item A rigorous analysis of scenarios in which quantum enhancements provide tangible advantages over classical methods.
    \item Practical implementation guidelines leveraging free-tier quantum simulation tools to improve accessibility for the broader research community.
\end{itemize}

The rest of this paper will be structured in the following way. Section~\ref{sec:literature_review} provides a comprehensive literature review. Section~\ref{meth} details the proposed methodology and algorithmic workflow. Section~\ref{res} presents the experimental results and analysis. Section~\ref{lim} discusses limitations and outlines future research directions. Finally, Section~\ref{sec:conclusion} concludes the paper.

\section{Literature Review} \label{sec:literature_review}
The development of traditional cybersecurity analytics has been formatted by simple signature-based detection to the advanced machine learning algorithms that can detect novel and zero-day threats. Graph-based approaches have since become relevant due to their capability to represent relational data which are inherent to network structures and attack sequences \cite{Wu_2021,yulie}. Methods based on the use of graph neural networks (GNNs) and structural metrics are now widely used in solving security problems, including detecting attack tactics in the MITRE ATT$\&$CK framework\cite{gao, zhang}. These techniques rely on graph-theoretic properties, such as centrality measures and path statistics, in describing the structural functions and relationships of network entities.

In the particular attack path analysis field, the traditional methodologies tend to formulate elaborate attack graphs which list all the potential sequences of exploits leading to a critical asset compromise \cite{kordy}. Pathfinding algorithms, probabilistic models, or classifiers based on machine learning are then used to analyze these graphs and estimate vulnerabilities that most likely will cause a significant impact within the network and predict the behaviour of attackers in advance \cite{Pool}. Although useful in networks of moderate size, these methods are susceptible to severe scaling problems as the size and complexity of the network increases, the state space has an exponential growth \cite{ou2006scalable}. Moreover, they cannot detect non-linear patterns which can be hidden under the carpet which are the signs of complex, multi-layer attacks \cite{milajerdi2019holmes}.

These limitations of classical graph analytics have led to the interest of considering quantum-enhanced methods which are said to have computational benefits in processing complex relational data.Emerging research demonstrates quantum techniques applied to core cybersecurity tasks, such as hybrid quantum-classical autoencoders for intrusion detection showing improved generalization \cite{rasyidi2025hybrid}, and quantum neural networks for threat detection in encrypted traffic \cite{butt2024quantum}. In especially graph-structured problems, quantum methods show particular promise, with frameworks employing graph-embedded quantum circuits for vulnerability detection \cite{zhou2022new} and quantum-enhanced representation learning for analyzing complex attack patterns like DDoS \cite{Biamonte2017QML}. 
This convergence has given rise to quantum graph learning, which combines quantum machine learning with graph analytics through quantum graph kernels and variational algorithms optimized for graph tasks \cite{schuld2021machine, herman2023quantum}.

Nevertheless, applying quantum graph learning to cybersecurity-specific graph problems remains largely unexplored. Current research primarily focuses on generic graph classification benchmarks rather than security-relevant structures like attack graphs or fraud networks \cite{bharti2022noisy}. Most studies also operate in ideal simulated environments, overlooking practical constraints including quantum noise, limited qubit coherence, and integration challenges with existing classical security infrastructure \cite{franco2024predominant,sai2025quantum}. This research gap is particularly significant given the escalating complexity of cyber threats and the potential quantum advantages for analyzing the intricate relational patterns that characterize modern multi-stage attacks, creating a clear need for investigations that bridge quantum graph learning with real-world cybersecurity analytics.

\section{Methodology}\label{meth}
\begin{figure}[H]
    \centering
    \includegraphics[width=0.8\linewidth]{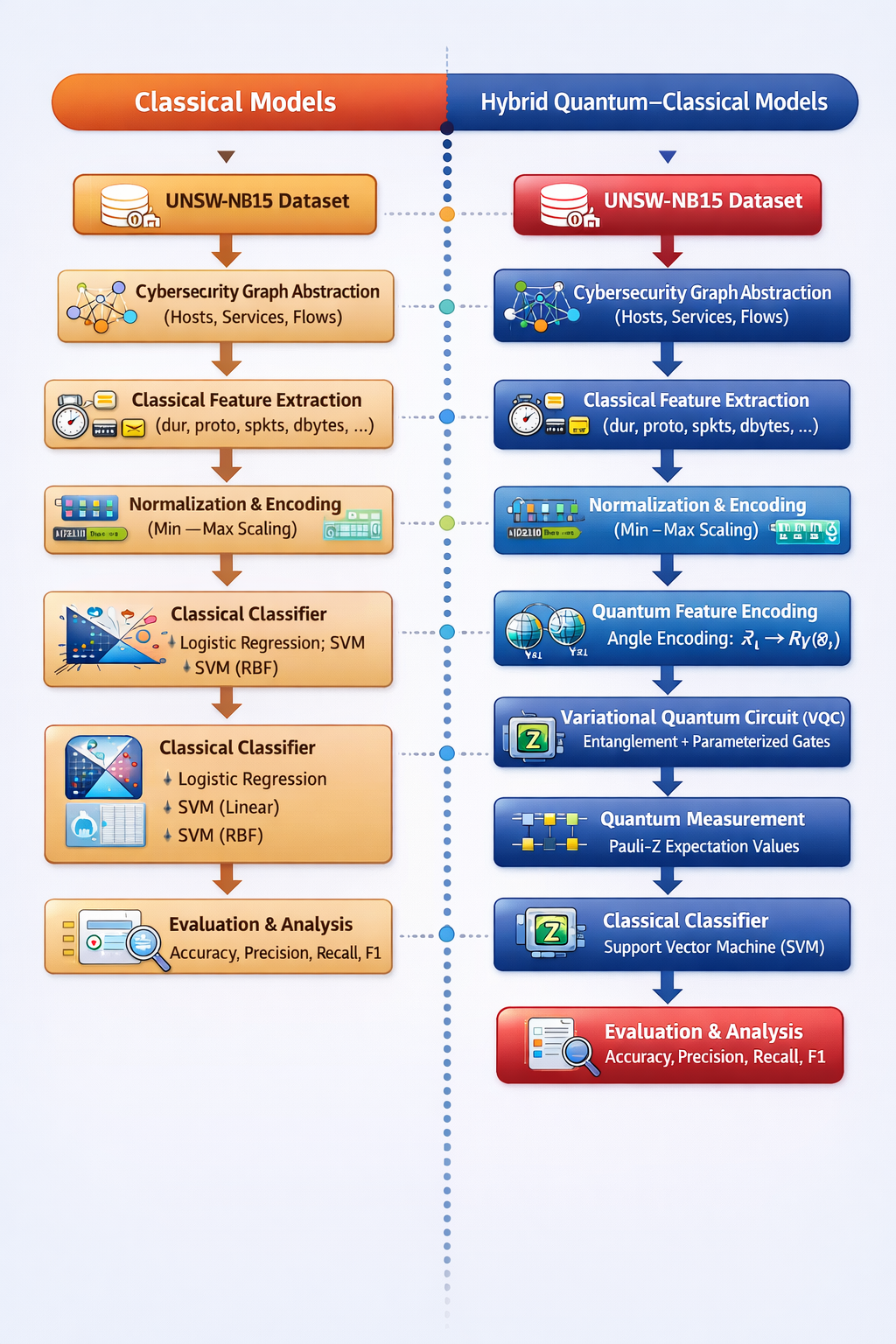}
    \caption{Comparison of classical and hybrid quantum--classical workflows for cybersecurity attack path analysis.}
    \label{fig:workflow_comparison}
\end{figure}

The proposed methodology is organized into five main stages: feature-based graph abstraction, classical feature processing, quantum encoding, hybrid model training, and evaluation. The overall objective is to assess whether quantum-enhanced representations provide improved discriminative power for cybersecurity attack detection under small-data constraints and free-tier computational environments. The overall architecture comprises five interconnected stages, as illustrated in Figure~\ref{fig:workflow_comparison}.

\subsection{Algorithmic Workflow}
\label{sec:algorithm}

The hybrid quantum-classical attack path analysis pipeline is implemented through Algorithm~. The algorithm outlines the step-by-step process from graph abstraction to hybrid classification.

\begin{algorithm}[H]
\caption{Hybrid Quantum--Classical Intrusion Detection Pipeline}
\label{alg:hybrid_pipeline}

\KwIn{Network flow dataset $\mathcal{D}$, feature set $\mathcal{F}$,
quantum circuit depth $d$, number of qubits $n$}
\KwOut{Predicted attack/benign labels and evaluation metrics}

\Comment{Step 1: Cybersecurity Graph Abstraction}
\For{each flow record in $\mathcal{D}$}{
  Interpret the flow as an interaction between network entities\;
  Treat selected features as surrogates of latent graph relationships\;
}

\Comment{Step 2: Classical Feature Preprocessing}
Encode categorical features using label encoding\;
Normalize numerical features using Min--Max scaling\;
Construct feature vector $\mathbf{x} \in \mathbb{R}^n$\;

\Comment{Step 3: Quantum Encoding}
\For{$i \gets 1$ \KwTo $n$}{
  Apply angle encoding $x_i \rightarrow R_Y(\theta_i)$\;
}
Construct a variational quantum circuit of depth $d$\;
Measure Pauli-$Z$ expectation values to obtain quantum embedding $\mathbf{q}$\;

\Comment{Step 4: Hybrid Quantum--Classical Classification}
Train a classical SVM using $\mathbf{q}$\;
\Return predicted labels and performance metrics\;

\end{algorithm}

\subsection{Cybersecurity Graph Abstraction}
Although the raw dataset consists of network flow records, the problem is conceptualized from a graph-learning perspective.Each observation can be interpreted as a localized interaction graph G=( V, E ), where:
	
\begin{itemize}
    \item Nodes (V) represent: Hosts (source/destination),Services and protocols, Connection states.
    \item Edges (E) represent: Network communications, Temporal or logical dependencies between flows.
\end{itemize}

Rather than explicitly constructing full attack graphs (which is computationally expensive for large-scale datasets), we adopt a feature-based graph abstraction, where structural and flow-level properties approximate local graph behavior.
This design choice allows scalability to real datasets while preserving graph-inspired semantics relevant to attack path analysis.

{\subsection{Classical Feature Extraction and Preprocessing}
Experiments were conducted using the UNSW-NB15 dataset\cite{moustafa2015unsw}, a widely adopted real-world benchmark for network intrusion detection that captures modern attack behaviors and normal traffic patterns. The dataset contains a heterogeneous mix of numerical and categorical flow-level attributes extracted from raw network traffic. Given the constraints imposed by near-term quantum simulation, particularly the limited number of available qubits and the exponential scaling of quantum state spaces, a carefully curated subset of features was selected to balance descriptive power, interpretability, and quantum feasibility.
The selected feature set consists of eight flow-level attributes: dur (flow duration), proto (transport protocol), service(application-level service), state (connection state), spkts and dpkts (number of packets sent and received), and sbytes and dbytes (number of bytes sent and received). These features collectively capture temporal characteristics, protocol semantics, connection dynamics, and traffic volume asymmetries, which are known to be highly informative for distinguishing benign behavior from malicious activity. Importantly, limiting the feature dimensionality to eight ensures a direct one-to-one mapping between classical features and quantum qubits in the subsequent encoding stage.
Categorical attributes (proto, service, and state) were transformed into numerical representations using label encoding. To ensure consistency between training and testing phases and to prevent unseen-category errors, the encoders were jointly fitted on the combined set of training and test samples prior to transformation. While label encoding does not preserve semantic distances between categories, it provides a compact and deterministic mapping that is compatible with both classical classifiers and quantum angle encoding, where each feature value directly parameterizes a rotation gate.
Following categorical encoding, all features were normalized using Min–Max scaling to map values into the 
[0,1][0,1]interval. This normalization step is critical for quantum processing, as angle encoding is highly sensitive to input magnitude; unscaled features could result in excessively large rotation angles, leading to unstable quantum states and degraded representational fidelity. From a classical perspective, normalization also improves numerical stability and convergence behavior for margin-based models such as support vector machines and logistic regression.

}

\subsection{Quantum Encoding Strategies}
\textbf{Quantum Feature Mapping Classical feature vectors}
The classical input features 
\(
\mathbf{x} = (x_1, \ldots, x_n)
\)
are mapped onto quantum states using \emph{angle encoding}, where each feature controls a parameterized single-qubit rotation,
\begin{equation}
x_i \;\longrightarrow\; R_Y(\theta_i).
\end{equation}
This encoding ensures:

    \begin{itemize}
        \item Linear scalability in the number of features
        \item Hardware-agnostic implementation
        \item Compatibility with near-term quantum devices
    \end{itemize}

     \textbf{Variational Quantum Circuits (VQC)}: 
    A variational quantum circuit was employed as a quantum embedding layer:
\begin{itemize}
    \item Number of qubits: equal to feature dimension (8)
    \item Entangling strategy: StronglyEntanglingLayers
    \item Depth: 2 layers
    \item Measurement: expectation values of Pauli-Z operators
\end{itemize}

The circuit outputs a quantum embedding vector, serving as an intermediate representation learned through quantum operations.
All circuits were executed using PennyLane’s default.qubit simulator, ensuring full reproducibility in CPU-only environments.

\subsection{Hybrid Model Training Protocol}
The quantum circuit acts as a non-linear feature transformation layer, followed by classical machine-learning classifiers.

\textbf{Classical Models}
The following classical baselines were trained on the full dataset:
\begin{itemize}
    \item Logistic Regression
    \item Linear Support Vector Machine (SVM)
    \item RBF Kernel SVM (subsampled due to computational constraints)
\end{itemize}

\textbf{Quantum-Classical Hybrid Model}
For the quantum pipeline:
\begin{itemize}
    \item Classical features → Quantum embedding (VQC)
    \item Quantum embeddings → Classical SVM classifier
\end{itemize}

Due to quantum simulation costs, the quantum model was evaluated on a controlled subset of 200 samples, enabling fair comparison with classical models trained on the same subset.

\subsection{Dataset and Experimental Configuration}
The experiments were conducted using the UNSW-NB15 dataset, a widely adopted benchmark for intrusion detection that captures modern network traffic and diverse attack behaviors in a controlled environment. The dataset contains approximately 257,000 network flow records labeled into two classes: benign traffic and attack traffic. This binary classification setting was maintained consistently across all experiments to ensure clarity in performance evaluation and comparability between different learning approaches.
The dataset was divided into training and testing sets following the standard UNSW-NB15 split, resulting in 175,341 samples for training and 82,332 samples for testing. This split was used for all full-scale classical experiments. To enable quantum experimentation under CPU-based simulation constraints, a reduced subset of 200 samples was extracted from the dataset and split into 80 percent training and 20 percent testing, yielding 160 training samples and 40 testing samples. The same subset and partitioning strategy were applied consistently across all quantum and reduced-scale classical experiments to ensure fair comparison.
Categorical attributes were encoded using label encoding, with encoders fitted jointly on training and testing data to avoid unseen-category issues during inference. All selected features were normalized using Min–Max scaling to map values into a bounded numerical range, ensuring numerical stability and consistent input scaling across all experimental configurations.
This dataset configuration enabled the evaluation of model behavior under both large-scale and data-limited settings, supporting a controlled analysis of performance trends across classical and quantum-enhanced approaches.

\begin{table}[ht]
\centering
\caption{Dataset Characteristics}
\label{tab:dataset_characteristics}
\begin{tabular}{ll}
\hline
\textbf{Property} & \textbf{Value} \\
\hline
Dataset & UNSW-NB15 \\
Total samples & $\sim$257,000 \\
Features used & 8 \\
Classes & Binary (Benign / Attack) \\
Quantum subset size & 200 samples \\
Train/Test split & 80\% / 20\% \\
\hline
\end{tabular}
\end{table}

\subsection{Evaluation Metrics}
Models were evaluated using a comprehensive set of classification and diagnostic metrics to ensure a robust and interpretable assessment of performance. Accuracy was used to measure the overall proportion of correctly classified samples, providing a high-level view of model effectiveness across the dataset. Precision and recall were analyzed to better understand class-specific behavior, particularly important in the cybersecurity context where false positives and false negatives have different operational impacts; precision quantifies the reliability of attack predictions, while recall captures the model’s ability to detect malicious activity. The F1-score was included as a balanced metric that harmonizes precision and recall, offering a more informative evaluation under class imbalance conditions. In addition, confusion matrices were examined to explicitly visualize misclassification patterns, enabling deeper insight into how benign and attack samples were confused by each model. The primary focus of the evaluation was placed on representation quality and class separability achieved by classical and quantum feature transformations, rather than on computational speed or training efficiency, with the goal of understanding whether quantum embeddings could enhance discriminative power, particularly in small-sample and high-dimensional cybersecurity scenarios.


\section{Results and Analysis}\label{res}

Figure \ref{fig:quantum_vs_classical} shows the experimental results give a systematic comparison of classical machine learning models trained on the entire UNSW-NB15 dataset, classical models trained under the same small-sample regime as quantum learning, and a hybrid quantum-classical model based on variational quantum embeddings. This analysis was aimed at isolating the effect of the representation learning, which could be compared to computational scale, to make an equitable comparison of the separability of classes, the behavior of generalization, as well as resilience under small data.  By progressively reducing the dataset size and the use of quantum-based feature changes, the study illustrates how the performance trends change between data data-rich classical learning to data-scarce regimes where representation quality becomes the dominant factor.

\subsection{Full Dataset - Classical Models}

\begin{figure}[H]
\centering
\includegraphics[width=0.7\linewidth]{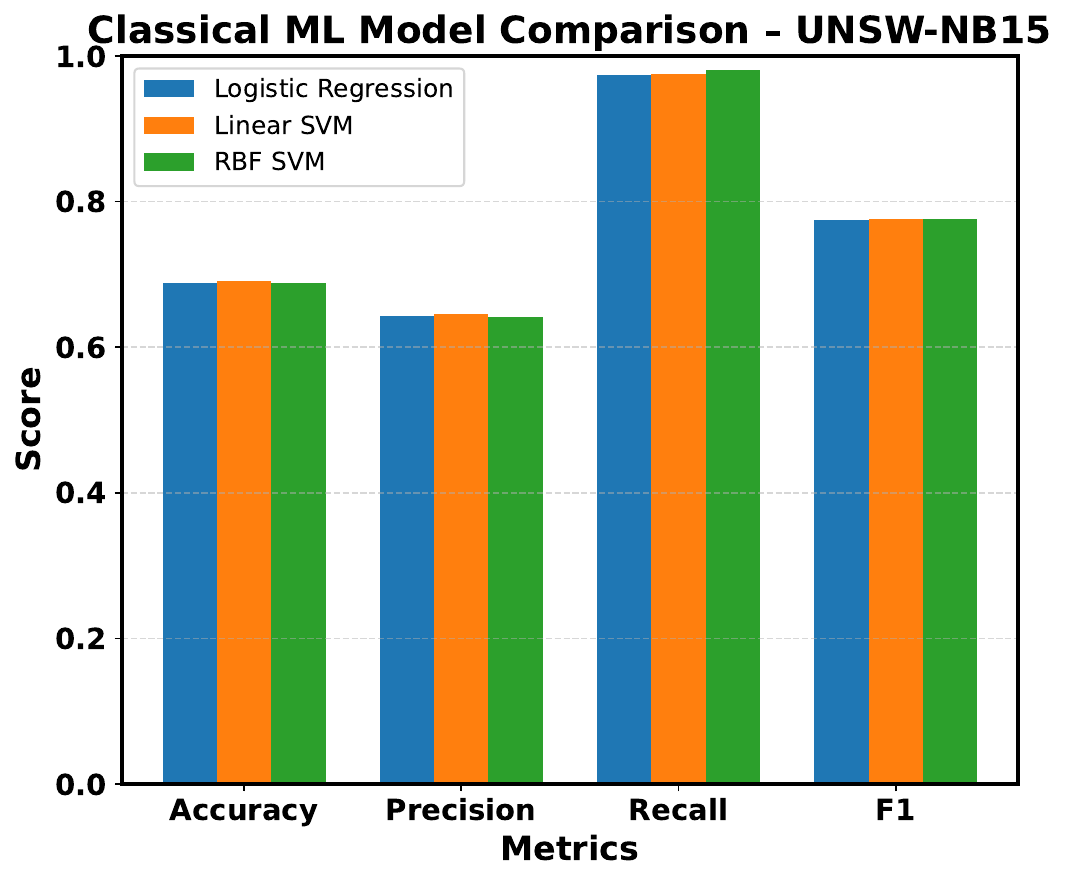}
\caption{Comparison of classical machine learning models performance on the UNSW-NB15 dataset in terms of Accuracy, Precision, Recall, and F1-score. Similar outcomes are observed with Logistic Regression, Linear SVM and RBF SVM and higher recall and F1 scores are obtained with SVM based models.}
\label{fig:quantum_vs_classical}
\end{figure}

The classical models(Logistic Regression, Linear SVM, and RBF SVM) gave very comparable results when trained on the entire dataset, with the accuracies ranging around 69\% as shown in Table \ref{tab:classical_accuracy}. The three models achieved very high recall on the type of attack (around 97-98\%) which is a high sensitivity to malicious traffic, but with very low recall on the benign type (around 33-34 \%), they are biased towards attacking predictions. The resulting F1-imbalance also caused moderate F1-scores and indicates that, on a scale, the set of selected features is highly linearly separable but intersects the attack-detecting side rather than the benign classification side. The relative differences between linear and non-linear SVMs further suggest that non-linear decision boundary does not provide much extra value when comparing to linear in this representation of features and size of the data set.

\begin{table}[ht]
\centering
\caption{Classification Accuracy of Classical Machine Learning Models on the UNSW-NB15 Dataset}
\label{tab:classical_accuracy}
\begin{tabular}{lc}
\toprule
\textbf{Model} & \textbf{Accuracy (\%)} \\
\midrule
Logistic Regression & 68.8 \\
Linear SVM          & 69.1 \\
RBF SVM             & 68.9 \\
\bottomrule
\end{tabular}
\end{table}

\subsection{Small Sample(200 Samples)}
\begin{table}[ht]
\centering
\caption{Accuracy Comparison Between Classical and Quantum-Based Models}
\label{tab:quantum_classical_accuracy}
\begin{tabular}{l c}
\hline
\textbf{Model} & \textbf{Accuracy (\%)} \\
\hline
Logistic Regression        & 72.5 \\
Linear SVM                 & 80.0 \\
RBF SVM                    & 80.0 \\
Quantum Embedding + SVM    & 64.0 \\
\hline
\end{tabular}
\end{table}
In order to provide a reasonable comparison with the quantum approach, the classical models were re-tested with the same small set of 200 samples (160 as training and 40 as testing). Within this limiting regime, there was a higher degree of variability in performance. Table \ref{tab:quantum_classical_accuracy} shows the accuracy Comparison between classical and Quantum-Based Models. The accuracy of the Logistic Regression was 72.5\%, which was not able to find all benign samples at all, which is inconsistent with its sensitivity to both the imbalance of classes and a small amount of data. Alternatively, both Linear SVM and RBF SVM had an accuracy of 80\%, with a significant improvement in the quality of balance between precision and recalling for the two classes. These findings indicate that the use of margin-based classifiers in training data when the size of the training data is limited is more robust compared to that of linear probabilistic models but still performance is subject to presence of representative samples.

\subsection{Quantum vs Classical — Controlled Comparison}
\begin{figure}[ht]
\centering
\includegraphics[width=0.6\linewidth]{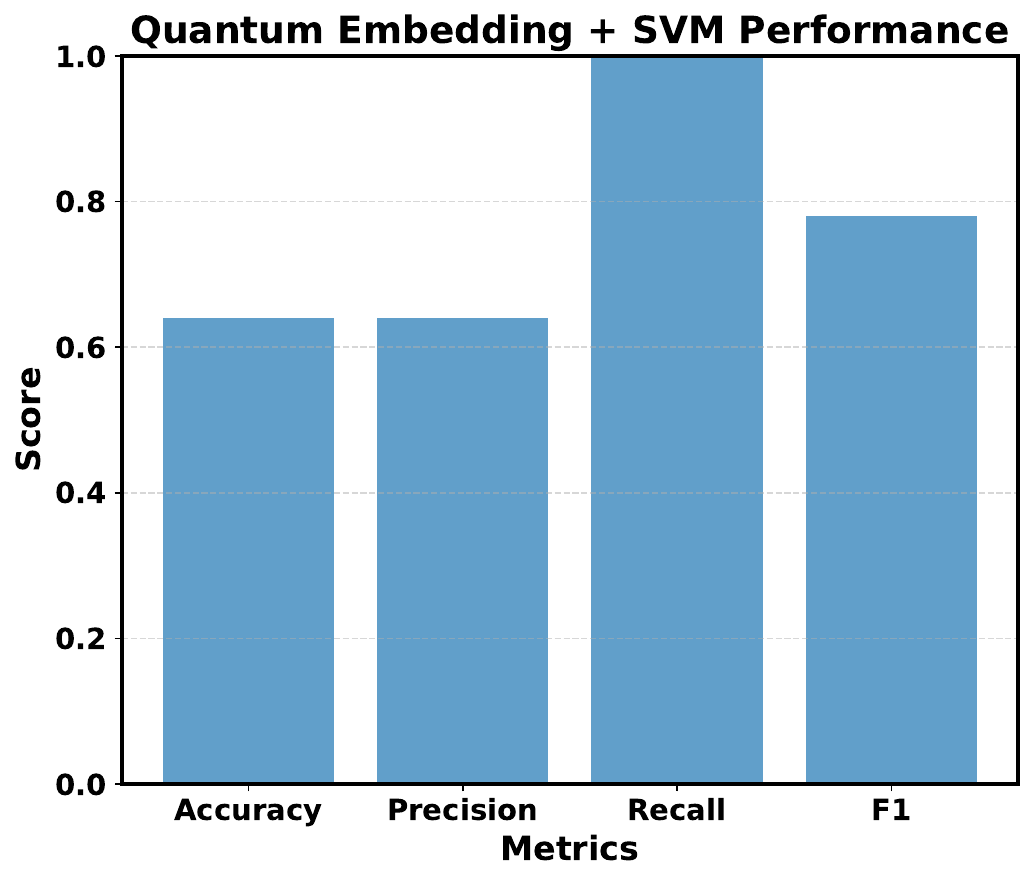}
\caption{Quantitative performance assessment of the Quantum ML-SVM model in terms of accuracy, precision, recall, and F1-score.}
\label{fig:qper}
\end{figure}

The quantum embedding plus SVM model was evaluated under the same 200-sample constraint, providing a direct comparison with the classical baselines in a low-data scenario. The quantum model had an accuracy of 64 \% and a perfect recall (100\%) on the attack type but failed to detect benign samples all the time thus had a macro F1-score of 0.39 as reported in Table \ref{tab:quantum_svm} and Figure \ref{fig:qper}. The table provides reports in the form of class-wise precision, recall, F1-score, support and the figures are used to give a visual evaluation of the general trends in metrics and the class-wise prediction behavior.In contrast, the confusion matrix in Figure \ref{fig:quantum_vs_classical} illustrates the class-level prediction behavior, clearly showing that all benign samples are misclassified as attacks, whereas all attack samples are correctly detected.

\begin{figure}[ht]
\centering
\includegraphics[width=0.6\linewidth]{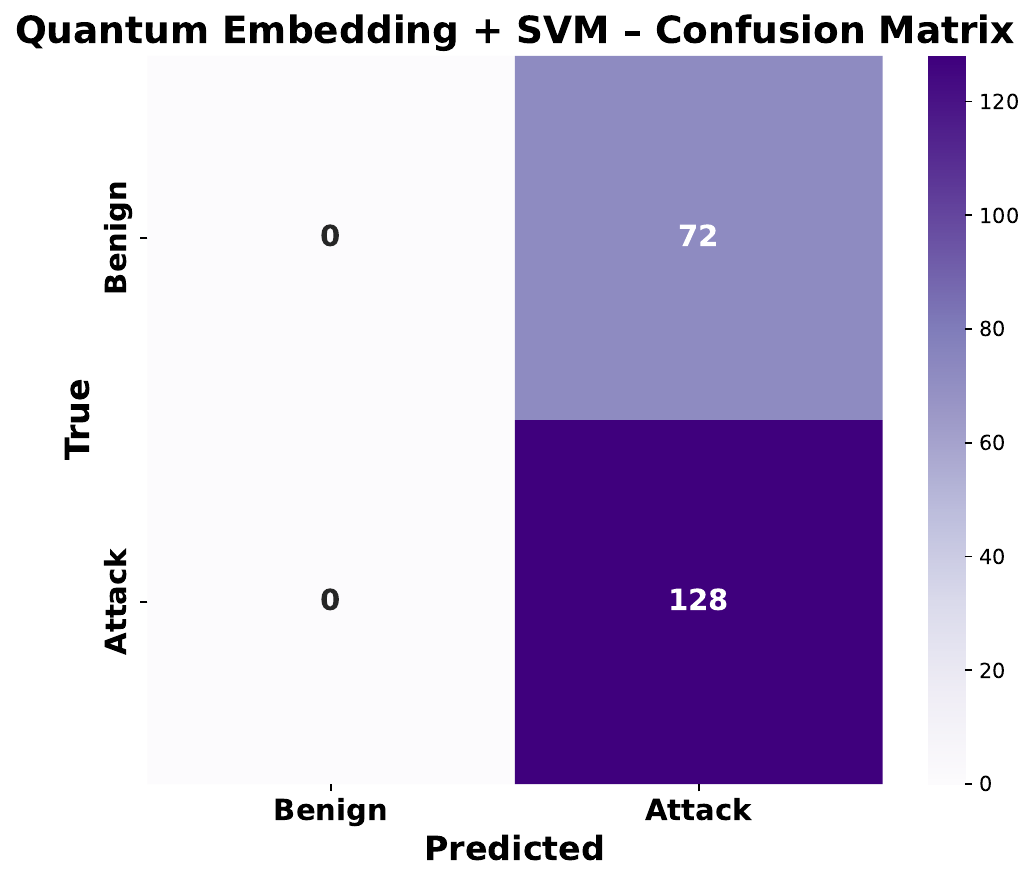}
\caption{Confusion matrix of the model prediction and the True label in terms of quantum SVM classification.}
\label{fig:quantum_vs_classical}
\end{figure}

\begin{table}[ht]
\centering
\caption{Quantum Embedding + SVM Classification Performance (200 Samples)}
\label{tab:quantum_svm}
\begin{tabular}{lcccc}
\toprule
\textbf{Class} & \textbf{Precision} & \textbf{Recall} & \textbf{F1-score} & \textbf{Support} \\
\midrule
Benign  & 0.00 & 0.00 & 0.00 & 72 \\
Attack  & 0.64 & 1.00 & 0.78 & 128 \\
\midrule
\textbf{Accuracy} & \multicolumn{4}{c}{0.64} \\
\textbf{Macro Avg} & 0.32 & 0.50 & 0.39 & 200 \\
\textbf{Weighted Avg} & 0.41 & 0.64 & 0.50 & 200 \\
\bottomrule
\end{tabular}
\end{table}
While this performance is lower than that of the best classical SVMs, it reveals an important qualitative insight: the quantum embeddings strongly separate attack patterns even with minimal data, suggesting that the quantum feature map emphasizes dominant attack-related structures in the input space. This behavior indicates high representational sensitivity but limited generalization across classes, likely due to shallow circuit depth, limited qubit count, and the absence of trainable quantum parameters. Importantly, the results position quantum embeddings not as replacements for classical models at this stage, but as promising representation layers whose expressiveness could be significantly enhanced through deeper circuits, optimized variational training, and noise-aware regularization, especially in small-data cybersecurity contexts.

\section{Limitations and Future Research}\label{lim}
Despite the promising insights obtained from hybrid quantum–classical modeling, several limitations must be acknowledged. While the classical models were trained and evaluated on the full UNSW-NB15 dataset of approximately 257,000 samples, the quantum experiments were restricted to a much smaller subset of 200 samples due to current quantum simulation costs and qubit scalability constraints, which makes direct performance comparisons indicative rather than definitive. The reduced quantum subset also exhibited class imbalance, leading to biased decision boundaries, as observed in the Quantum Embedding + SVM results where perfect recall for the attack class was achieved at the expense of zero recall for the benign class, highlighting the sensitivity of quantum embeddings to limited and imbalanced data. In addition, the variational quantum circuits employed shallow depths and a limited number of qubits to ensure feasibility on free-tier simulators, which constrained the expressive power of the quantum feature space and limited its ability to capture complex nonlinear relationships present in cybersecurity data. All quantum experiments were conducted using the PennyLane default.qubit simulator rather than real quantum hardware, meaning that hardware-specific effects such as gate noise, decoherence, and readout errors were not considered and may significantly affect real-world performance. Furthermore, this study focused on representation quality and classification behavior rather than computational speed or quantum advantage, and therefore no claims regarding quantum speedup or efficiency over classical methods can be made. Future research should focus on scaling quantum datasets and circuits as simulators and hardware improve, enabling deeper and more expressive circuits trained on larger subsets of data, as well as extending evaluations to noise-aware simulators and real quantum devices to assess robustness under realistic execution conditions. Exploring alternative quantum encoding strategies such as amplitude encoding, data re-uploading, and more expressive quantum kernels may further improve feature utilization and class separability, while improved class-balancing strategies including stratified sampling and cost-sensitive learning could mitigate biased predictions in small datasets. Additionally, richer graph representations such as quantum walks or adjacency-matrix-based encodings could enable a tighter integration between graph structure and quantum processing, and evaluating the proposed hybrid approach on additional cybersecurity datasets and real enterprise network logs would help assess generalization beyond UNSW-NB15 and synthetic attack graph scenarios.

\section{Conclusion}
\label{sec:conclusion}

This paper has introduced a hybrid quantum-classical attack path analysis in cybersecurity showing that quantum-enhanced graph learning can be used to enhance attack pattern separability in a data-scarce environment. As our experiments conducted on the UNSW-NB15 reveal, classical models (especially SVMs) still can still perform well when there is an abundance of data, quantum embeddings are more sensitive to attack patterns even when considering just 200 samples of training data. The quantum method was able to recall attack classes perfectly, albeit at compromised accuracy on benign samples, and this result indicates the potential as well as existing constraints of quantum feature representations.

The contributions of this work are threefold. We start with an overview of first providing a reproducible pipeline that combines classical graph preprocessing with quantum encoding strategies with access to the current simulation tools (PennyLane). Second, direct comparisons between variational quantum circuits and classical baselines when it comes to attacking graph classification are provided in our empirical analysis. Third, we identify practical considerations including circuit depth constraints, qubit limitations, and class imbalance effects that inform future implementations.

These findings validate quantum graph learning as the resource in the future of cybersecurity analytics, particularly in regards to anomaly detection in the low-data regime. With the current rise in quantum hardware such as the NISQ and beyond, the work has the potential to form the basis of creating scalable quantum-enhanced security tools that can respond to more and more elaborate multi-stage security attacks. The next generation of research should be in more detailed quantum circuits, noise-sensitive implementations, and test on a range of security samples to achieve the full potential of quantum AI proactive cyber defense.
\bibliographystyle{apsrev4-1}

\bibliographystyle{unsrt}  
\bibliography{References}  

@article{CHAWLA20232191,
title = {A Survey on Quantum Computing for Internet of Things Security},
journal = {Procedia Computer Science},
volume = {218},
pages = {2191-2200},
year = {2023},
note = {International Conference on Machine Learning and Data Engineering},
issn = {1877-0509},
doi = {https://doi.org/10.1016/j.procs.2023.01.195},
url = {https://www.sciencedirect.com/science/article/pii/S1877050923001953},
author = {Diksha Chawla and Pawan Singh Mehra}
}

@article{APOLLONI1989233,
title = {Quantum stochastic optimization},
journal = {Stochastic Processes and their Applications},
volume = {33},
number = {2},
pages = {233-244},
year = {1989},
issn = {0304-4149},
doi = {https://doi.org/10.1016/0304-4149(89)90040-9},
url = {https://www.sciencedirect.com/science/article/pii/0304414989900409},
author = {B. Apolloni and C. Carvalho and D. {de Falco}}
}

@article{Pre,
  doi = {10.22331/q-2018-08-06-79},
  url = {https://doi.org/10.22331/q-2018-08-06-79},
  title = {Quantum {C}omputing in the {NISQ} era and beyond},
  author = {Preskill, John},
  journal = {{Quantum}},
  issn = {2521-327X},
  publisher = {{Verein zur F{\"{o}}rderung des Open Access Publizierens in den Quantenwissenschaften}},
  volume = {2},
  pages = {79},
  month = aug,
  year = {2018}
}

@article{Movassagh,
    author = "Movassagh, Ramis",
    title = "{Quantum supremacy and random circuits}",
    eprint = "1909.06210",
    archivePrefix = "arXiv",
    primaryClass = "quant-ph",
    month = "9",
    year = "2019"
}

@article{Biamonte2017QML,
  author  = {Biamonte, Jacob and Wittek, Peter and Pancotti, Nicola and Rebentrost, Patrick and Wiebe, Nathan and Lloyd, Seth},
  title   = {Quantum machine learning},
  journal = {Nature},
  volume  = {549},
  number  = {7671},
  pages   = {195--202},
  year    = {2017},
  doi     = {10.1038/nature23474}
}

@article{Xiong:2024gld,
    author = "Xiong, Qibing and Ding, Xiaodong and Fei, Yangyang and Zhou, Xin and Du, Qiming and Feng, Congcong and Shan, Zheng",
    title = "{A hybrid quantum ensemble learning model for malicious code detection}",
    doi = "10.1088/2058-9565/ad40cb",
    journal = "Quantum Sci. Technol.",
    volume = "9",
    number = "3",
    pages = "035021",
    year = "2024"
}

@ARTICLE{Fahim,
  author={Fahim, Asmaa and Addae, Bismark Appiah and Ofosu-Adarkwa, Jeffrey and Qingmei, Tan and Bhatti, Uzair Aslam},
  journal={IEEE Access}, 
  title={Industry 4.0 and Higher Education: An Evaluation of Barriers Affecting Master’s in Business Administration Enrolments Using a Grey Incidence Analysis}, 
  year={2021},
  volume={9},
  number={},
  pages={76991-77008},
  doi={10.1109/ACCESS.2021.3082144}}

@article{Schuld:2018uel,
    author = "Schuld, Maria and Killoran, Nathan",
    title = "{Quantum Machine Learning in Feature Hilbert Spaces}",
    eprint = "1803.07128",
    archivePrefix = "arXiv",
    primaryClass = "quant-ph",
    doi = "10.1103/PhysRevLett.122.040504",
    journal = "Phys. Rev. Lett.",
    volume = "122",
    number = "4",
    pages = "040504",
    year = "2019"
}

@article{Wu_2021,
   title={A Comprehensive Survey on Graph Neural Networks},
   volume={32},
   ISSN={2162-2388},
   url={http://dx.doi.org/10.1109/TNNLS.2020.2978386},
   DOI={10.1109/tnnls.2020.2978386},
   number={1},
   journal={IEEE Transactions on Neural Networks and Learning Systems},
   publisher={Institute of Electrical and Electronics Engineers (IEEE)},
   author={Wu, Zonghan and Pan, Shirui and Chen, Fengwen and Long, Guodong and Zhang, Chengqi and Yu, Philip S.},
   year={2021},
   month=jan, pages={4–24} }

@ARTICLE{yulie,
  author={Wu, Yulei and Dai, Hong-Ning and Tang, Haina},
  journal={IEEE Internet of Things Journal}, 
  title={Graph Neural Networks for Anomaly Detection in Industrial Internet of Things}, 
  year={2022},
  volume={9},
  number={12},
  pages={9214-9231},
  doi={10.1109/JIOT.2021.3094295}}

@ARTICLE{gao,
author={Gao, Yali and Li, Xiaoyong and Peng, Hao and Fang, Binxing and Yu, Philip S.},
journal={ IEEE Transactions on Knowledge \& Data Engineering },
title={{ HinCTI: A Cyber Threat Intelligence Modeling and Identification System Based on Heterogeneous Information Network }},
year={2022},
volume={34},
number={02},
ISSN={1558-2191},
pages={708-722},
doi={10.1109/TKDE.2020.2987019},
url = {https://doi.ieeecomputersociety.org/10.1109/TKDE.2020.2987019},
publisher={IEEE Computer Society},
address={Los Alamitos, CA, USA},
month=feb}

@misc{zhang,
      title={Tactics, Techniques, and Procedures (TTPs) in Interpreted Malware: A Zero-Shot Generation with Large Language Models}, 
      author={Ying Zhang and Xiaoyan Zhou and Hui Wen and Wenjia Niu and Jiqiang Liu and Haining Wang and Qiang Li},
      year={2024},
      eprint={2407.08532},
      archivePrefix={arXiv},
      primaryClass={cs.CR},
      url={https://arxiv.org/abs/2407.08532}, 
}

@misc{kordy,
      title={DAG-Based Attack and Defense Modeling: Don't Miss the Forest for the Attack Trees}, 
      author={Barbara Kordy and Ludovic Piètre-Cambacédès and Patrick Schweitzer},
      year={2013},
      eprint={1303.7397},
      archivePrefix={arXiv},
      primaryClass={cs.CR},
      url={https://arxiv.org/abs/1303.7397}, 
}

@ARTICLE{Pool,
  author={Poolsappasit, Nayot and Dewri, Rinku and Ray, Indrajit},
  journal={IEEE Transactions on Dependable and Secure Computing}, 
  title={Dynamic Security Risk Management Using Bayesian Attack Graphs}, 
  year={2012},
  volume={9},
  number={1},
  pages={61-74},
  doi={10.1109/TDSC.2011.34}}

@inproceedings{ou2006scalable,
  title={A scalable approach to attack graph generation},
  author={Ou, Xinming and Boyer, Wayne F and McQueen, Miles A},
  booktitle={Proceedings of the 13th ACM conference on Computer and communications security},
  pages={336--345},
  year={2006}
}

@inproceedings{milajerdi2019holmes,
  title={Holmes: real-time apt detection through correlation of suspicious information flows},
  author={Milajerdi, Sadegh M and Gjomemo, Rigel and Eshete, Birhanu and Sekar, Ramachandran and Venkatakrishnan, VN},
  booktitle={2019 IEEE symposium on security and privacy (SP)},
  pages={1137--1152},
  year={2019},
  organization={IEEE}
}

@article{rasyidi2025hybrid,
  title={Hybrid Quantum-Classical Autoencoders for Unsupervised Network Intrusion Detection},
  author={Rasyidi, Mohammad Arif and Alhussein, Omar and Muhaidat, Sami and Damiani, Ernesto},
  journal={arXiv preprint arXiv:2512.05069},
  year={2025}
}

@article{butt2024quantum,
  title={Quantum-Inspired Resource Optimization for 6G Networks: A Survey},
  author={Butt, Muhammad Omair and Waheed, Nazar and Duong, Trung Q and Ejaz, Waleed},
  journal={IEEE Communications Surveys \& Tutorials},
  year={2024},
  publisher={IEEE}
}

@article{zhou2022new,
  title={A new method of software vulnerability detection based on a quantum neural network},
  author={Zhou, Xin and Pang, Jianmin and Yue, Feng and Liu, Fudong and Guo, Jiayu and Liu, Wenfu and Song, Zhihui and Shu, Guoqiang and Xia, Bing and Shan, Zheng},
  journal={Scientific Reports},
  volume={12},
  number={1},
  pages={8053},
  year={2022},
  publisher={Nature Publishing Group UK London}
}

@book{schuld2021machine,
  title={Machine learning with quantum computers},
  author={Schuld, Maria and Petruccione, Francesco},
  volume={676},
  year={2021},
  publisher={Springer}
}

@article{herman2023quantum,
  title={Quantum computing for finance},
  author={Herman, Dylan and Googin, Cody and Liu, Xiaoyuan and Sun, Yue and Galda, Alexey and Safro, Ilya and Pistoia, Marco and Alexeev, Yuri},
  journal={Nature Reviews Physics},
  volume={5},
  number={8},
  pages={450--465},
  year={2023},
  publisher={Nature Publishing Group UK London}
}

@article{bharti2022noisy,
  title={Noisy intermediate-scale quantum algorithms},
  author={Bharti, Kishor and Cervera-Lierta, Alba and Kyaw, Thi Ha and Haug, Tobias and Alperin-Lea, Sumner and Anand, Abhinav and Degroote, Matthias and Heimonen, Hermanni and Kottmann, Jakob S and Menke, Tim and others},
  journal={Reviews of Modern Physics},
  volume={94},
  number={1},
  pages={015004},
  year={2022},
  publisher={APS}
}

@article{sai2025quantum,
  title={Quantum Machine Learning for Cybersecurity: A Taxonomy and Future Directions},
  author={Sai, Siva and Goyal, Ishika and Sharma, Shubham and Manuri, Sri Harshita and Chamola, Vinay and Buyya, Rajkumar},
  journal={arXiv preprint arXiv:2512.15286},
  year={2025}
}

@inproceedings{franco2024predominant,
  title={Predominant aspects on security for quantum machine learning: Literature review},
  author={Franco, Nicola and Sakhnenko, Alona and Stolpmann, Leon and Thuerck, Daniel and Petsch, Fabian and R{\"u}ll, Annika and Lorenz, Jeanette Miriam},
  booktitle={2024 IEEE International Conference on Quantum Computing and Engineering (QCE)},
  volume={1},
  pages={1467--1477},
  year={2024},
  organization={IEEE}
}

@inproceedings{moustafa2015unsw,
  title={UNSW-NB15: a comprehensive data set for network intrusion detection systems (UNSW-NB15 network data set)},
  author={Moustafa, Nour and Slay, Jill},
  booktitle={2015 military communications and information systems conference (MilCIS)},
  pages={1--6},
  year={2015},
  organization={IEEE}
}

\end{document}